\documentclass{article}

\usepackage{microtype}
\usepackage{graphicx}
\usepackage{subfigure}
\usepackage{booktabs} 
\usepackage{tabularx}
\usepackage{listings}

\usepackage{url}
\usepackage{hyperref}

\usepackage[accepted]{icml2025}

\usepackage{amsmath}
\usepackage{amssymb}
\usepackage{mathtools}
\usepackage{amsthm}

\usepackage[capitalize,noabbrev]{cleveref}

\theoremstyle{plain}

\theoremstyle{definition}

\theoremstyle{remark}

\usepackage[textsize=tiny]{todonotes}

\begin{document}

\twocolumn[
\icmltitle{Construction and Evaluation of LLM-based agents for Semi-Autonomous penetration testing}



\icmlsetsymbol{equal}{*}

\begin{icmlauthorlist}
\icmlauthor{Masaya Kobayashi}{meiji}
\icmlauthor{Masane Fuchi}{meiji}
\icmlauthor{Amar Zanashir}{lac}
\icmlauthor{Tomonori Yoneda}{lac}
\icmlauthor{Tomohiro Takagi}{meiji}
\end{icmlauthorlist}

\icmlaffiliation{meiji}{Meiji University, Kanagawa, Japan}
\icmlaffiliation{lac}{LAC Co. Ltd., Tokyo, Japan}

\icmlcorrespondingauthor{Masaya Kobayashi}{msy.kbysh02@gmail.com}

\icmlkeywords{LLM-based Agent, Penetration Testing}
\vskip 0.3in
]



\printAffiliationsAndNotice{}  

\begin{abstract}
With the emergence of high-performance large language models (LLMs) such as GPT, Claude, and Gemini, the autonomous and semi-autonomous execution of tasks has significantly advanced across various domains. However, in highly specialized fields such as cybersecurity, full autonomy remains a challenge. This difficulty primarily stems from the limitations of LLMs in reasoning capabilities and domain-specific knowledge. We propose a system that semi-autonomously executes complex cybersecurity workflows by employing multiple LLMs modules to formulate attack strategies, generate commands, and analyze results, thereby addressing the aforementioned challenges. In our experiments using Hack The Box virtual machines, we confirmed that our system can autonomously construct attack strategies, issue appropriate commands, and automate certain processes, thereby reducing the need for manual intervention.
\end{abstract}

\section{Introduction}
The real-world applications of large language models (LLMs) are increasing, with coding assistance and business process autonomy being representative examples. LLMs possess the ability to execute complex tasks on the basis of text-based instructions, and their potential has been demonstrated in various application scenarios ~\cite{wu2023autogenenablingnextgenllm, wang2023voyageropenendedembodiedagent, chan2023chatevalbetterllmbasedevaluators, Wang_2024}.

In the field of cybersecurity, LLMs have been applied to tasks such as vulnerability and phishing detection, autonomy and diversification of attacks, and information manipulation ~\cite{Yao_2024}. Notably, dynamic and flexible approaches using LLMs have demonstrated performance comparable with or exceeding traditional static methods in areas such as vulnerability discovery through fuzzing, code remediation, and the creation of targeted phishing emails.~\cite{noever2023largelanguagemodelsfix, koide2025detectingphishingsitesusing, Xia_2024} However, in attack simulations and vulnerability assessments of operating systems, LLMs face challenges in adapting to environments and acquiring and utilizing specialized knowledge, limiting their effectiveness.

We focus on penetration testing and explore methods for leveraging LLMs to achieve semi-autonomy. Research on LLM-assisted penetration testing includes expert support systems for strategy formulation and execution planning ~\cite{DBLP:conf/uss/DengLVLLX0R0024}, autonomy focused on post-exploitation ~\cite{xu2024autoattacker}, privilege escalation ~\cite{Happe_2023, happe2024llmshackersautonomouslinux}, web-based attacks ~\cite{fang2024llmagentsautonomouslyhack}, zero-day attack autonomy ~\cite{fang2024llmagentsautonomouslyexploit}, and one-day attack autonomy ~\cite{fang2024teamsllmagentsexploit}. However, these studies are limited in terms of the attack phases they automate. Research that attempts to automate the entire penetration testing process ~\cite{huang2023penheal} relies on specific environments and tools, which impose numerous operational constraints, making general application in real-world scenarios challenging.

Considering real-world applications, a less restrictive system is required. Therefore, we propose a novel system for autonomous penetration testing. Inspired by recent advancements in prompt engineering, such as Self-Refine ~\cite{madaan2023selfrefine} and ReAct ~\cite{yao2023reactsynergizingreasoningacting}, we design an LLM-based agent architecture and integrate information retrieval techniques, such as retrieval-augmented generation (RAG) ~\cite{NEURIPS2020_6b493230}, to efficiently utilize specialized knowledge for strategy formulation and execution planning.

In our experiments using Hack The Box~\cite{noauthor_hack_nodate} virtual machines, the results demonstrate that our system can autonomously construct attack strategies, generate appropriate commands, and automate certain processes, significantly reducing the need for manual intervention. However, since most existing approaches are not publicly available, direct comparisons are not possible, and quantitative evaluation using benchmarks has yet to be conducted. Thus, the evaluation of our system remains qualitative. Nevertheless, experimental results suggest that our system overcomes the limitations of conventional methods and offers a flexible, autonomous penetration testing framework capable of adapting to diverse attack scenarios.
\section{Backgrounds}
\subsection{Penetration Testing}
Penetration testing is a method for assessing system vulnerabilities from the perspective of an actual attacker, traditionally conducted manually by security experts. In recent years, various autonomous approaches utilizing LLMs have been proposed, including support systems ~\cite{DBLP:conf/uss/DengLVLLX0R0024}, autonomous systems focusing on specific attack stages ~\cite{xu2024autoattacker, Happe_2023}, systems for automating vulnerability exploitation ~\cite{fang2024llmagentsautonomouslyexploit, fang2024teamsllmagentsexploit}, and frameworks that attempt to autmate the entire penetration testing process ~\cite{huang2023penheal}. These approaches aim to autonomously generate and execute attack scenarios, which is difficult with traditional methods. However, a common challenge among them is that they have not yet achieved full autonomy.
\subsection{Retrieval-augmented Generation}
Retrieval-augmented generation~\cite{NEURIPS2020_6b493230, gao2024retrievalaugmentedgenerationlargelanguage} is a technique that dynamically incorporates external knowledge during text generation, overcoming the limitation of traditional LLMs relying solely on knowledge acquired during training. A representative method, Hybrid Search ~\cite{lukawski_hybrid_nodate}, integrates keyword-based search with semantic vector search to enhance information retrieval accuracy. Additionally, Reranking ~\cite{noauthor_reranking_nodate} is used to select the most relevant information from retrieved data. These methods enable the generation of responses that reflect the latest information and specialized domain knowledge.
\subsection{Prompt Engineering}
Prompt engineering is a technique for designing input prompts to optimize LLMs' output. Chain-of-Thought prompting ~\cite{wei2023chainofthoughtpromptingelicitsreasoning} facilitates step-by-step reasoning for complex problems by leveraging internal knowledge. On the other hand, Self-Refine ~\cite{madaan2023selfrefine} enables an LLM to generate feedback on its initial output and iteratively refine it on the basis of that feedback. Additionally, ReAct ~\cite{yao2023reactsynergizingreasoningacting} is a prompting method that alternates between reasoning and action generation, enabling the model to dynamically incorporate external information while planning and executing high-level tasks. These techniques enhance the reasoning capabilities of LLMs in complex security tasks without requiring additional training.
\subsection{LLM-based Agents}
LLM-based agents leverage the powerful generative capabilities of LLMs to autonomously perform decision-making, planning, interaction, and integration with external environments beyond simple text generation. Traditional LLMs have been widely used as static tools that generate responses on the basis of given inputs~\cite{kojima2023largelanguagemodelszeroshot, brown2020languagemodelsfewshotlearners}. However, LLM-based agents extend this capability to handle complex problem-solving and interactive tasks~\cite{wu2023autogenenablingnextgenllm, wang2023voyageropenendedembodiedagent, chan2023chatevalbetterllmbasedevaluators}. By integrating internal reasoning with external actions, LLM-based agents enable more transparent and sophisticated decision-making processes.
\section{Proposed Method}
\begin{figure}[tbp]
  \centering
  \includegraphics[width=0.8\linewidth]{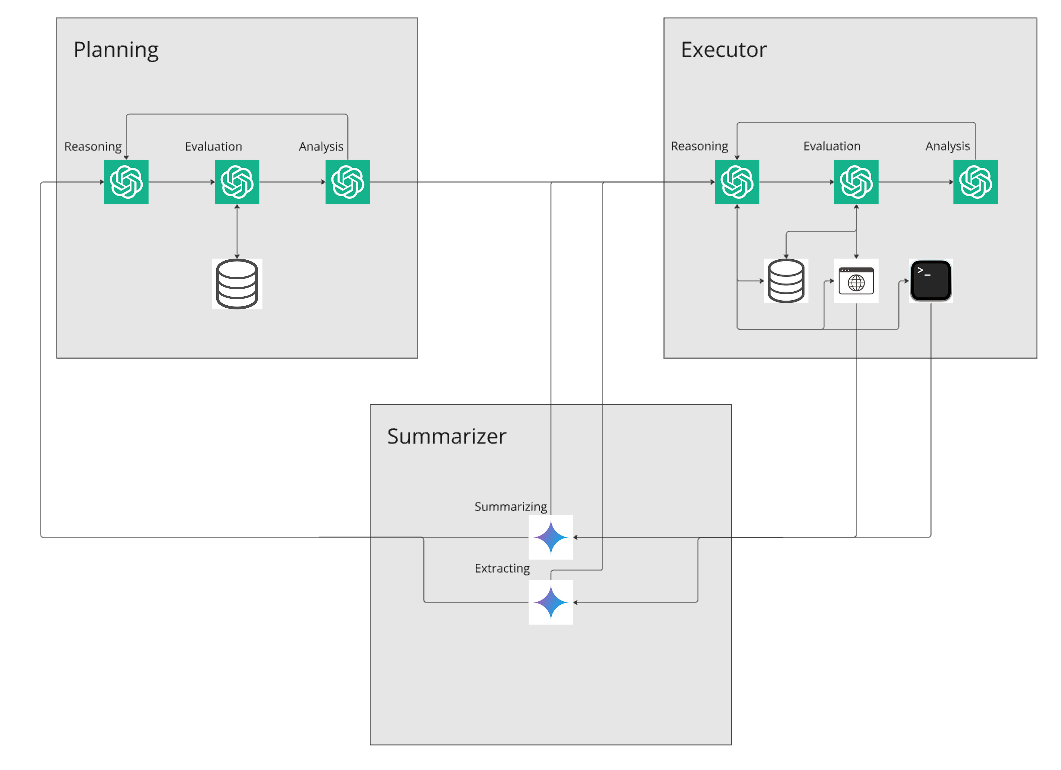}
  \caption{Overview of our proposed method.}
  \label{figure1}
\end{figure}
\subsection{Overview}
Our semi-autonomous penetration testing system consists of multiple LLMs, each assigned distinct roles in reasoning, evaluation, and analysis. These LLMs are organized into three modules: planning, execution, and summarization. Additionally, the system integrates RAG, search engines, and an execution environment as tools. Figure 1 provides an overview. In this system, human intervention is minimized to decision-making and monitoring of command execution. The system is designed to be independent of specific environments, enabling penetration testing using tools executable on the Kali Linux command line.
\subsection{Planning Module}
The planning module formulates attack strategies for penetration testing and manages the entire intrusion process. It employs the Pentesting Task Tree (PTT), a data structure used in related research, to track attack phases. On the basis of the summarization module's results, the system updates the state and information of tasks and subtasks for each attack phase. The PTT structure in this study is built upon the MITRE ATT\&CK framework ~\cite{noauthor_mitre_nodate} and ensures that all recorded execution results are accessible to every module.

The module's architecture draws inspiration from Self-Refine and ReAct. The roles of PTT generation and updating (reasoning), feedback provision on inference results (evaluation), and information retrieval and improvement proposal generation (analysis) are distributed among multiple LLMs. This iterative process refines the PTT. Furthermore, by incorporating execution results, the system ensures the generation of accurate execution procedures.
\subsection{Execution Module}
The execution module translates attack strategies into concrete execution steps and generates terminal-executable commands. Similar to the planning module, commands undergo refinement through the reasoning, evaluation, and analysis processes of multiple LLMs. Human operators review the generated commands and explanations to determine whether to execute them in the terminal while monitoring the process. This intervention is necessary because LLMs remain susceptible to hallucinations, and some operations—such as those requiring root privileges or actions that may compromise system integrity—demand human oversight.
\subsection{Summarization Module}
The summarization module consists of LLMs dedicated to summarization and extraction. The summaries of execution and search results are strictly on the basis of retrieved information, excluding any speculative reasoning or predictions from the LLMs. The extraction process identifies and isolates critical data, such as passwords and authentication credentials. This ensures that even if the execution results contain redundant information, only essential details are passed on to the planning and execution modules, improving efficiency and clarity in decision-making.
\subsection{Tools}
\subsubsection{Information Gathering}
For LLMs to effectively leverage specialized knowledge, efficient information gathering is essential. We employ a combination of RAG and web searches to collect relevant data. When the agent determines that additional information is necessary, it selects an appropriate tool and generates a search query accordingly.

For RAG, useful penetration testing resources such as MITRE ~\cite{noauthor_mitre_nodate}, OWASP ~\cite{noauthor_owasp_nodate}, and references for Kali Linux Tools ~\cite{noauthor_get_nodate} are stored in a Qdrant ~\cite{noauthor_qdrant_nodate} vector database. Relevant documents are retrieved on demand.

For web searches, the system can extract information from sources such as the National Vulnerability Database (NVD) ~\cite{nvd} and publicly available repositories on GitHub ~\cite{github}. This approach provides access to the latest security knowledge.
\subsubsection{Execution Environment}
The execution environment enables multiple terminal operations using subprocess (Python) and tmux (Linux). On the basis of execution procedures generated by the execution module, specified commands are executed in designated terminal sessions.
\section{Experiment}
\subsection{Experimental Environment}
We conducted experiments on virtual machines provided by Hack The Box. The target machine was selected on the basis of the criteria that the official Hack The Box difficulty rating was ``Easy'' and the user rating was more than 4.5, with a preference for machines incorporating a relatively large number of technical elements. On the basis of these criteria, we selected Board Light\footnote{https://www.hackthebox.com/machines/boardlight} as the test target. Table 1 presents the testing methods used in our experiment.

For our proposed method, we used GPT-4o ~\cite{openai2024gpt4ocard} for the planning and execution modules, while Gemini-1.5-flash ~\cite{geminiteam2024gemini15unlockingmultimodal} was utilized for the summarization module. Additionally, for the vector database, text-embedding-3-large ~\cite{openai_vector_embeddings} was used as the embedding model, and Rerank 3 ~\cite{cohere_rerank} was employed as the reranking model.
\begin{table}[tbp] 
\caption{Overview of Board Light} 
\begin{center} 
\begin{tabularx}{\linewidth}{X|X} 
\toprule 
Element & Description \\
\midrule 
Port Scanning & Port scanning using Nmap \\ 
\midrule
Web Application Analysis & Information gathering using WhatWeb and Curl \\
& Subdomain and directory enumeration using Ffuf and Gobuster \\ 
\midrule
Initial Exploitation & Exploit utilizing CVE-2023-30253 \\
\midrule
Authentication & Privilege escalation to another user on the basis of authentication credentials \\
\midrule
Privilege Escalation & Information gathering using Linpeas \\
& Exploit utilizing CVE-2022-37706 \\
\bottomrule
\end{tabularx}
\end{center}
\end{table}
\subsection{Results and Discussion}
Table 2 presents a subset of the tasks and commands generated during the test (only relevant sections are extracted). Our system successfully performed an autonomous penetration testing, except for the part involving directory and file exploration within the server during a password attack.

We found that in the planning module, initial inference results sometimes failed to generate the necessary strategies. However, through evaluation and analysis, the system was able to refine and correct these strategies. This demonstrates that LLMs that engage in iterative reasoning cycles are effective in deepening their insights. Table 3 compares the initial and refined generations of the planning module. In this table, the initial plan only included an attack strategy targeting SSH (port 22), whereas after evaluation, an additional strategy for HTTP (port 80) was also generated.

In the execution module, tasks that could be executed with a single command were mostly generated correctly on the first attempt through reasoning and analysis. Table 4 compares the initial and refined generations of the execution module. The initial command generation did not correctly utilize user credentials or passwords, but after refinement, the execution steps were corrected on the basis of errors and retrieved information. However, when multiple commands were required such as retrieving and executing an exploit script from GitHub,— errors occurred several times. To address this, the system dynamically executed web searches on the basis of the generated queries, retrieved references such as README files, and refined the commands accordingly. This highlights the effectiveness of tools that supplement specialized knowledge, such as Kali Linux references. This ability to utilize tools and process large amounts of information is a key strength of LLM-based agents, enabling autonomous penetration testing for known vulnerabilities.

However, when conducting post-exploitation tasks such as exploring directories and files within the compromised server, although the system selected appropriate tools, it demonstrated poor sensitivity to environment-specific details such as directory names and file names. The extensive trial-and-error process required for this exploration made task completion more challenging. Furthermore, unlike other tools, this process relies on internal system information rather than external sources, highlighting a limitation of information retrieval tools that do not guarantee accurate data. This also suggests the limitations of applying a generalized LLMs to every stage of a penetration testing.

To address these challenges in the future, we believe that adopting a knowledge graph instead of a vector database for information retention and creating specialized agents tailored to each attack phase would be effective. Assigning dedicated agents to specific tasks would facilitate the development of a more advanced LLM-based agent system.
\begin{table}[tbp]
\caption{Planning and executing module results}
\begin{center}
\scalebox{0.8}{
\begin{tabularx}{\linewidth}{X|X}
\toprule
Generated task & Generated command \\
\midrule
Perform a port scan & nmap -sS -sV 10.10.11.11 \\
\midrule
Web Information Gathering & 
\lstinline |whatweb -a 3 -v http://10.10.11| \newline
\lstinline |curl -I http://10.10.11.11| \\
\midrule
Subdomain enumeration for ``board.htb'' & 
\lstinline |gobuster dns -d board.htb -w bitquark-subdomains-top100000.txt| \newline
\lstinline |ffuf -w subdomains-top1million-5000.txt -u http://10.10.11.11 -H "Host:FUZZ.crm.board.htb"| \\
\midrule
Web technology stack identification for subdomains & whatweb -a 3 http://crm.board.htb \\
\midrule
Search for CVEs and exploit codes on NVD and GitHub & 
\lstinline | search: Dolibarr version17.0.0 CVE | \newline
\lstinline | search: site:github.com Dolibarr 17.0.0 exploit | \newline
\lstinline | search: site:nvd.nist.gov Dolibarr 17.0.0 CVE | \\
\midrule
Exploit remote command execution and file upload vulnerabilities in Dolibarr & 
\lstinline | git clone https://github.com/nikn0laty/Exploit-for-Dolibarr-17.0.0-CVE-2023-30253.git | \newline
\lstinline | cd Exploit-for-Dolibarr-17.0.0-CVE-2023-30253 | \newline
\lstinline | nc -lnvp 4444 | \newline
\lstinline | python3 exploit.py | \newline
\lstinline | crm.board.htb admin | \newline
\lstinline | admin 10.10.14.2 4444 | \\
\midrule
Search for configuration files & 
\lstinline | find / -type f ( -name "*.conf" -o -name "*.ini" -o -name "*.env" -o -name "*.yaml" -o -name "*.php | \\
\midrule
Identify password files and sensitive information & 
\lstinline | cat /etc/password | \newline
\lstinline | find / -perm -u=s -type f 2>/dev/null | \\
\midrule
Attempt SSH connections with discovered credentials &  
\lstinline | sshpass -p `serverfun2\$2023!!' ssh larissa@10.10.11.11 -p 22 | \\
\midrule
Search for additional vulnerabilities & 
\lstinline | enlightment -version | \\
\midrule
Exploit system services for privilege escalation & 
\lstinline | search: enlightment 0.23.1 CVE | \newline
\lstinline | git clone https://github.com/MaherAzzouzi/CVE-2022-37706-LPE-exploit | \newline
\lstinline | cd CVE-2022-37706-LPE-exploit | \newline
\lstinline | scp exploit.sh larissa@10.10.11.11:exploit.sh | \newline
\lstinline | ./exploit.sh | \\
\bottomrule
\end{tabularx}
}
\end{center}
\end{table}

\begin{table}[tbp]
\caption{Capabilities of the planning module (partial list)}
\begin{center}
\scalebox{0.9}{
\begin{tabularx}{\linewidth}{X|X}
\toprule
Initial Generation & Modification \\
\midrule
\lstinline|1 Reconnaissance - [completed]| \newline
\lstinline|...| \newline
\lstinline|1.2 Identify Open Ports and Services - (completed)| \newline
\lstinline|1.2.1 Perform a port scan - (completed)| \newline
\lstinline|- result:| \newline
\lstinline|- **Open Ports**: 22/tcp (ssh), 80/tcp (http)| \newline
\lstinline|- **Services**:| \newline
\lstinline|- 22/tcp: ssh OpenSSH 8.2p1 Ubuntu 4ubuntu0.11 (Ubuntu Linux; protocol 2.0)| \newline
\lstinline|- 80/tcp: http Apache httpd 2.4.41 ((Ubuntu))| \newline
\lstinline|- **OS**: Linux (detected by nmap)| \newline
\lstinline|- **Network Distance**: 2 hops| \newline
\lstinline|- **SSH Host Keys**:| \newline
\lstinline|- RSA: 3072 ...| \newline
\lstinline|- ECDSA: ...| \newline
\lstinline|- ED25519: 256 ...| \newline
\lstinline|1.2.2 Detailed scan on specific ports - (completed)| \newline
\lstinline|- result:| \newline
\lstinline|- **Service Info**: OS: Linux; CPE: cpe:/o:linux:linux\_kernel| \newline
\lstinline|- **HTTP Server Header**: Apache/2.4.41 (Ubuntu)| \newline
\lstinline|- **HTTP Title**: Site doesn't have a title (text/html; charset=UTF-8).| \newline
\lstinline|2 Initial Access - [to-do]| \newline
\lstinline|2.1 Exploit Discovered Services - (to-do)| \newline
\lstinline|2.1.1 Attempt SSH Access - (to-do)| \newline
\lstinline|2.1.2 Investigate HTTP Service for Vulnerabilities - (to-do)| \newline
&
\lstinline|1 Reconnaissance - [completed]| \newline
\lstinline|...| \newline
\lstinline|1.3 Web Information Gathering - [to-do]| \newline
\lstinline|1.3.1 Identify web technologies and frameworks - (to-do)| \newline
\lstinline|1.3.2 Enumerate web directories and files - (to-do)| \newline
\lstinline|1.3.3 Analyze web content for sensitive information or misconfigurations - (to-do)| \newline
\lstinline|2 Initial Access - [to-do]| \newline
\lstinline|...| \newline
\lstinline|2.2 Vulnerability Assessment - [to-do]| \newline
\lstinline|2.2.1 Check for known CVEs related to Apache and OpenSSH - (to-do)| \newline
\lstinline|2.2.2 Investigate potential misconfigurations or outdated software versions - (to-do)| \\
\bottomrule
\end{tabularx}
}
\end{center}
\end{table}

\begin{table}[tbp]
\caption{Capabilities of the execution module (selected excerpts)}
\begin{center}
\scalebox{0.9}{
\begin{tabularx}{\linewidth}{X|X}
\toprule
Initial Generation & Modification \\
\midrule
\lstinline|git clone https://github.com/...| \newline
\lstinline|cd ...| \newline
\lstinline|python3 exploit.py http://crm.board.htb <USERNAME> <PASSWORD> 10.10.14.2 4444| \newline
&
\lstinline|git clone https://github.com/...| \newline
\lstinline|cd ...| \newline
\lstinline|python3 exploit.py http://crm.board.htb admin admin 10.10.14.2 4444| \\
\bottomrule
\end{tabularx}
}
\end{center}
\end{table}

\section{Conclusion and Future Work}
We proposed an LLM-based agent system utilizing multiple LLM modules to address the challenges of applying LLMs to autonomous penetration testing, specifically the lack of specialized knowledge and the complexity of reasoning. Through experiments, we confirmed that the system can autonomously conduct penetration testing by leveraging complex reasoning across multiple LLMs and flexible information retrieval. However, we also identified limitations, particularly in exploratory approaches to environment-specific configurations.

Additionally, due to the difficulty of enabling LLMs to operate the GUI of penetration testing tools, we constrained our implementation to command-line executable tools. However, by integrating systems such as Anthropic's Computer Use ~\cite{noauthor_computer_nodate} and OpenAI's Operator ~\cite{noauthor_operator_nodate}, it may be possible to develop a more advanced LLM-based agent system capable of executing even more sophisticated penetration testing tasks.

We plan to continue refining our approach by addressing the limitations identified in this study and improving the overall system architecture.

\bibliography{references}
\bibliographystyle{icml2025}

\end{document}